\documentclass[conference]{IEEEtran}
\IEEEoverridecommandlockouts
\usepackage{cite}
\usepackage{amsmath,amssymb,amsfonts}
\usepackage{algorithmic}
\usepackage{multirow}
\usepackage{graphicx}
\usepackage{textcomp}
\usepackage{listings}
\usepackage[table,xcdraw]{xcolor}
\usepackage[caption=false]{subfig}
\usepackage{url}
\usepackage{hyperref}
\usepackage{tikz}
\usepackage{dblfloatfix}

\newcommand\rounded[2]{
    \protect\tikz[baseline=(char.base)]{
        \protect\node[draw = #1, fill = white, rectangle, rounded corners, 
              text = black](char){#2};
  }
}%

\definecolor{codegreen}{rgb}{0,0.6,0}
\definecolor{codegray}{rgb}{0.5,0.5,0.5}
\definecolor{codepurple}{rgb}{0.58,0,0.82}
\definecolor{backcolour}{rgb}{0.95,0.95,0.92}
\definecolor{codeblue}{rgb}{0.13, 0.29, 0.53}
\definecolor{codeorange}{rgb}{0.81, 0.36, 0.0}

\definecolor{clr-background}{RGB}{255,255,255}
\definecolor{clr-text}{RGB}{0,0,0}
\definecolor{clr-string}{RGB}{163,21,21}
\definecolor{clr-namespace}{RGB}{0,0,0}
\definecolor{clr-preprocessor}{RGB}{128,128,128}
\definecolor{clr-keyword}{RGB}{0,0,255}
\definecolor{clr-type}{RGB}{43,145,175}
\definecolor{clr-variable}{RGB}{0,0,0}
\definecolor{clr-constant}{RGB}{111,0,138} 
\definecolor{clr-comment}{RGB}{0,128,0}
\definecolor{roundseagreen}{RGB}{0, 255, 255}
\definecolor{roundblue}{RGB}{0, 127, 255}
\definecolor{roundgreen}{RGB}{0, 204, 0}
\definecolor{roundorange}{RGB}{255, 179, 102}

\lstdefinestyle{mystyle}{
	backgroundcolor=\color{backcolour},
	commentstyle=\color{clr-comment},
    keywordstyle=\color{codeblue}\textbf,
 	stringstyle=\color{clr-string},
  	identifierstyle=\color{clr-variable}, 
    stringstyle=\color{codepurple},
    basicstyle=\ttfamily\footnotesize,
    breakatwhitespace=false,         
    breaklines=true,                 
    captionpos=t,                    
    keepspaces=true,                 
    numbers=left,                    
    numbersep=4pt,                  
    showspaces=false,                
    showstringspaces=false,
    showtabs=false,                  
    tabsize=2
}

\lstdefinelanguage{chisel}{
  language     = scala,
  morekeywords = [2]{Input, Output, IO, Wire, Reg, RegInit},
  morekeywords = [3]{UInt, SInt, Bool, Clock, Reg, Vec, Bundle, ChiselEnum, Module},
  morekeywords = [4]{},
  keywordstyle = [2]\color{codepurple},
  keywordstyle = [3]\color{codeorange},
  keywordstyle = [4]\textbf,
  sensitive = true,
}

\def\BibTeX{{\rm B\kern-.05em{\sc i\kern-.025em b}\kern-.08em
    T\kern-.1667em\lower.7ex\hbox{E}\kern-.125emX}}
\begin{document}

\title{Tywaves: A Typed Waveform Viewer for Chisel
}

\author{\IEEEauthorblockN{Raffaele Meloni}
\IEEEauthorblockA{\textit{Delft University of Technology}\\
Delft, The Netherlands \\
raffaele.meloni99@gmail.com
}
\and
\IEEEauthorblockN{H. Peter Hofstee}
\IEEEauthorblockA{\textit{IBM Infrastructure} (Austin, TX, US)\\
\textit{Delft University of Technology} (NL)\\
hofstee@us.ibm.com}
\and
\IEEEauthorblockN{Zaid Al-Ars}
\IEEEauthorblockA{\textit{Delft University of Technology}\\
Delft, The Netherlands \\
Z.Al-Ars@tudelft.nl}

}

\maketitle

\begin{abstract}
Chisel (Constructing Hardware In a Scala Embedded Language) is a broadly adopted HDL that brings object-oriented and functional programming, type-safety, and parameterization to hardware design. 
However, while these language features significantly improve the process of writing code, debugging Chisel designs with open source tools loses many of the advantages of the source language, as type information and data structure hierarchies are lost in the translation, simulator output, and waveform viewer.
This work, Tywaves, presents a new type-centered debugging format that brings the same level of abstraction found in contemporary hardware languages to waveform viewers. 
Contributions to the Chisel library and CIRCT MLIR compiler as well as the Surfer waveform viewer result in a waveform viewer that better supports the Chisel HDL. 
%

\noindent 
Project url: https://github.com/rameloni/tywaves-chisel-demo
\end{abstract}


\section{Introduction}
\label{sec:intro}


The end of Moore’s law slowed the rate of improvement of general-purpose processors, which led to increased research towards new specialized processors and accelerators~\cite{hennessyNewGoldenAge2019journalArticlea}.
Hardware development for a specialized design typically begins with a software version of an algorithm that is subsequently adapted to a hardware description language~(HDL) that defines the functionality of a digital design. 
However, classic~HDLs (e.g., Verilog and VHDL) are known for requiring significant design effort due to their low-level nature~\cite{shachamRethinkingDigitalDesign2011}.
In response, the current hardware design domain has experienced an explosion of modern~HDLs and hardware generator languages (HGLs) that introduce new levels of abstraction to make hardware design easier~\cite{truongGoldenAgeHardware2019journalArticle}. 
Chisel~\cite{bachrachChiselConstructingHardware2012conferencePaper}, Spade~\cite{skarmanSpadeHDLInspired2022conferencePaper} and Clash~\cite{baaijASHStructuralDescriptions2010conferencePaper} are examples of new modern hardware languages that incorporate software programming features such as complex type systems, object-oriented and functional programming, to facilitate and speed up the development. 

Despite this progress, less work has been done on improving the debugging infrastructure. 
To keep compatibility with pre-existing designs and industry tools (i.e., synthesis, verification, implementation, simulation, etc.) these languages are usually translated into classic~HDLs during the process of hardware generation. 
Therefore, hardware debugging tools must work on automatically generated code that is dissimilar from the source, leading to a debugging experience that does not retain the advantages of the source language.
Waveform viewers are an example of widely used tools affected by this problem where their graphical representation of signals does not reflect what developers specify in their original high-level code.



Tywaves is a type-centered waveform viewer for the Chisel hardware language that integrates its functionality within the Chisel Scala library and the CIRCT MLIR compiler  compiler~\cite{lenharthCIRCTLiftingHardware2021journalArticle} that is used to generate Verilog.
The viewer is based on the Tydi (Typed dataflow interface) ecosystem, consisting of the Tydi spec~\cite{tydi}, Tydi language (Tydi-lang)~\cite{tydilang} and Tydi-Chisel~\cite{tydichisel}. 
However, the contributions described in this paper can also be used in a broader context without dependency on the Tydi ecosystem.
This work aims to reduce the gap in abstraction between the source code and waveforms by keeping the code structure and showing high-level signal types. 
In addition to its graphical interface, Tywaves implements a methodology for propagating generic debug information from a source language throughout the CIRCT compiler. 
Furthermore, it adopts a modular architecture that facilitates future extensions. 
The specific contributions of this paper are: 
\begin{enumerate}
    \item Updates to the Chisel and CIRCT repositories to extract and link source-level data types with the compiled output;
    \item Adding support for Chisel and CIRCT within an existing waveform viewer;
    \item Creating a Rust library, \texttt{tywaves-rs}, that processes and converts debug information into a more generic and efficient data structure for the UI;
    \item An API to easily use the updated tools with the new functionality.
\end{enumerate}

The paper is organized as follows. Section~\ref{sec:background} provides a background description of Chisel, CIRCT, and typed hardware in the Chisel language. 
Section~\ref{sec:related-work} presents related work on debugging tools for Chisel. 
Contributions to the Chisel and CIRCT compilation pipeline, integration of Tywaves into an existing waveform viewer, and an API for easy adoption into existing Chisel simulations are discussed in Section~\ref{sec:implementation}. 
In~Section~\ref{sec:use-case}, the output results are shown by testing an example CPU implementing OpenPOWER ISA~\cite{OpenPOWERFoundationwebpage}.
Section~\ref{sec:summary} concludes the paper.

\section{Background}
\label{sec:background}
This section provides a general background on Chisel, CIRCT compilation, and Chisel code simulation.
We also present and describe typed circuit components in Chisel using a concrete example.

\subsection{Chisel}
Chisel~\cite{bachrachChiselConstructingHardware2012conferencePaper} is a broadly adopted HDL that brings object-oriented and functional programming, type-safety, and parameterization to hardware design. In addition, Chisel provides a strong type system for complex data to further reduce the effort needed for defining large and sophisticated designs.
Moreover, since it is embedded in the Scala programming language, it supports the implementation and sharing of libraries similarly to software applications; thus, improving code reusability in the hardware domain.

\subsection{Chisel compilation and CIRCT}
\begin{figure}
    \centering
    \includegraphics[width=\linewidth]{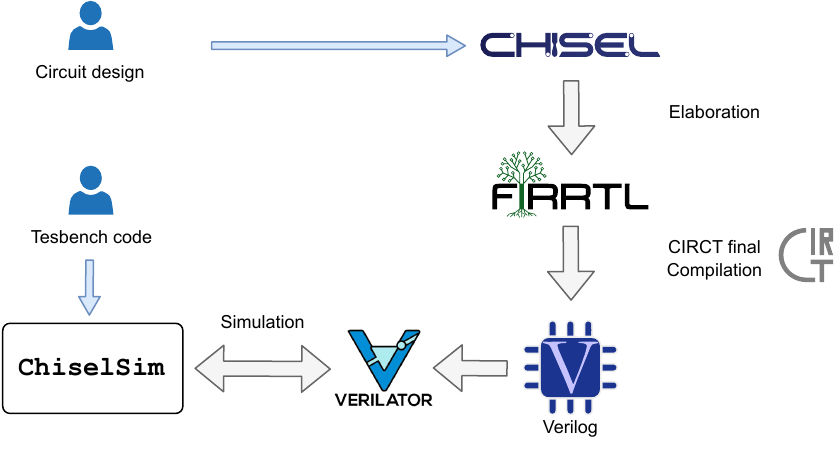}
    \caption{Chisel compilation and simulation flow}
    \label{fig:chisel-comp-sim-flow}
\end{figure}

As mentioned, Chisel, like other modern languages, is transformed into a classic HDL during the hardware generation process. Specifically, it is compiled into Verilog in two steps as depicted by Figure~\ref{fig:chisel-comp-sim-flow}:
\begin{enumerate}
    \item \textbf{Chisel elaboration}: the Chisel code is translated into an equivalent FIRRTL circuit. FIRRTL~\cite{izraelevitzReusabilityFIRRTLGround2017conferencePaper} (Flexible Internal Representation for RTL) is an intermediate representation (IR) that is used to decouple the Verilog transformation from the high-level Scala code used to describe Chisel circuits. This decoupling reduces the complexity of the compilation process~\cite{chowIntermediateRepresentationIncreasing2013journalArticle}.
    \item \textbf{CIRCT compilation}: FIRRTL is parsed, optimized, and compiled to Verilog.
\end{enumerate}

CIRCT~\cite{eldridge2021mlir} is an open source project that implements an~HDL compiler by applying the MLIR~\cite{lattnerMLIRScalingCompiler2021conferencePaper} and LLVM~\cite{lattnerLLVMCompilationFramework2004conferencePapera} development methodologies to the domain of hardware design tools, including a shared and reusable compiling infrastructure with interoperability between multiple languages, enabling language-independent transformations and optimizations.

The core components of CIRCT are represented by the MLIR dialects: unique namespaces to abstract and manipulate languages and IRs within a compiler. 
Dialects group and model multiple characteristics of an IR with Operations (Ops). An operation outputs zero or more results, can receive multiple input operands, and may have multiple attributes to represent data at runtime and compile-time.
Multiple Ops can be combined to represent data, control flow, and other properties of a language using static single assignments~\cite{cytronEfficientlyComputingStatic1991journalArticle}, which specify that a result from an Op can be assigned only once to an operand of another Op.

As noted earlier, CIRCT specifically targets Chisel through FIRRTL which is consequently parsed and converted into its corresponding MLIR dialect. 
This representation is lowered to simplify complex constructs and transformed into language-independent IRs, called CIRCT core dialects that perform the final transformations to emit the outputs from the compiler. 

\subsection{Chisel simulation}
Although simulating the output of Verilog is certainly possible, writing a testbench in such a low-level language would not be convenient in the context of Chisel, since all the abstractions introduced by the language would be lost in the test code. 
In addition, testbench compatibility would not be guaranteed because the compiled code may change unexpectedly when the source is updated. 
Therefore, the Chisel library provides a high-level simulation component to write tests and run simulations directly from Scala, called ChiselSim.

As shown in Figure~\ref{fig:chisel-comp-sim-flow}, the ChiselSim component implements a bridging interface between the user testbench code and the low-level Verilog simulator. 
Moreover, ChiselSim can be integrated with ScalaTest to improve test organization and simplify test execution with \texttt{sbt} or more advanced IDE interfaces.

\subsection{CIRCT debug dialect}
The debug dialect is a core dialect that is responsible for tracking the relationship of values, types, and hierarchies of an input language in CIRCT to fully reconstruct its representation from the target compiled output~\cite{DebugDialectCIRCTwebpage} that might have flattened or even optimized out variables.
It is composed of 4 MLIR~Ops that can be combined to represent and reconstruct the original hierarchical view: 
\begin{itemize}
    \item \textit{dbg.variable}: represents a named variable declared in the source code.
    \item \textit{dbg.struct} and \textit{dbg.array}: create aggregates from lists of named and indexed values, respectively. These Ops preserve the hierarchical structure of the variable independently from the optimizations that might be performed. 
    \item \textit{dbg.scope}: define a scope in the source code. It creates a namespace to group variables and other scopes, including module instances. 
\end{itemize}

The debug dialect is independent of the input language, as it is part of the core dialects, and it can be generated from any other MLIR dialect, including FIRRTL. 
When used within the compiler, it is serialized and emitted as an open standard format called HGLDD (Hardware Generator Language Debug Data) based on JSON. 
This file can be combined by a waveform viewer with the output trace file from a low-level RTL simulator (i.e., VCD) to reconstruct the source view. 

Unfortunately, the current state of CIRCT only allows the reconstruction of FIRRTL from Verilog rather than from a higher-level language such as Chisel.

\subsection{Typed hardware}
So far we have described what Chisel is and how it is transformed into Verilog through the CIRCT compiler. We also briefly mentioned that Chisel provides a type-system for complex data to bridge the gap between hardware and software development. 

A data type in a software language defines what kind of values a variable can have, e.g., \texttt{char} can have only character values while \texttt{int} represents signed integers within a specific range, although they are both represented as numbers in assembly or bytes in machine code.
Therefore, when developers inspect variables from a debugger, they expect to see the high-level value associated with the actual type of variables.

Likewise, elements of a Chisel circuit, such as signals, temporary variables, or module instances, are declared as Scala variables of a certain high-level type.
Basic and aggregate types can be used to associate one or multiple values to a single variable or to create enum-like types to represent named values.
Furthermore, developers can combine these types to create a type name and add parameters for custom data type definitions as shown in Listing~\ref{lst:types-chisel}. 
These type definitions can be reused multiple times for declaring ports, wires, registers, or module instances.
Table~\ref{tab:types-chisel} reports the types in Listing~\ref{lst:types-chisel} and the kind of associated values. 
Although FIRRTL preserves the code hierarchies it does not retain the Chisel Scala types (e.g., \texttt{MyState} is represented as a FIRRTL \texttt{UInt<2>}).

\begin{lstlisting}[language=chisel, caption=Example of Chisel types, label=lst:types-chisel]
// Struct-like with nested types
class MyBundle(val n: Int) extends Bundle {
  // Ground types  
  val a = Bool()
  val b = SInt(n.W)
  // Aggregates
  val s = new Bundle { val x = UInt(8.W) } 
  val v = Vec(n, SInt(32.W)) 
} 
// Enum-like type
class MyState extends ChiselEnum { 
  val IDLE, A, B, C, Other = Value 
} 
// User-defined module 
class MyModule(val n1: Int, val n2: Int, val n3: Int) extends Module { 
    val inBundle   = IO(Input(new MyBundle(n=n1))) 
    val wireBundle = Wire(new MyBundle(n2)) 
    val outBundle  = IO(Output(new MyBundle(n3))) 
    val state      = RegInit(MyState.IDLE) 
    // ... 
} 

class TopCircuit extends Module { 
    val mod1 = Module(new MyModule(10, 7, 9)) 
    val mod2 = Module(new MyModule(1, 1, 1)) 
    // ... 
}
\end{lstlisting}

\begin{table}[t]
    \centering
    \caption{Example of Chisel types and associated definitions and values}
    \label{tab:types-chisel}
    \begin{tabular}{lll}
        \hline
        \textbf{Type} & \textbf{Definition} & \textbf{Values} \\ \hline
        \texttt{IO[Bool]} & Port & \{True; False\} \\
        \texttt{Wire[SInt<10>]} & Wire signal & [$-2^9; +2^9-1$] \\
        \texttt{Reg[MyBundle]} & Register & A user defined-aggregate type \\
        \texttt{IO[SInt<32>[10]]} & Port & Indexed 32-bit signed integers \\
        \texttt{Reg[MyState]} & Register & \texttt{IDLE, A, B, C, Other}
        \\\hline
    \end{tabular}
\end{table}

\section{Related work}
\label{sec:related-work}
In the context of hardware development, we can distinguish between tools for testing and tools for debugging.
Testing involves the simulation of designs and verification of whether components operate as intended, while debugging is the process of identifying and removing any errors. 
Both techniques can enhance and contribute to the general process of hardware development with new tools that make the process faster and more user-friendly.

Recent developments have introduced three new debugging tools that are particularly relevant to Tywaves.
The Hardware Generator Debugger~\cite{zhangBringingSourcelevelDebugging2022} is a novel debugging tool that enables breakpoint debugging for inspecting signal value changes within the same clock cycle, helping to identify possible errors that may occur during intermediate calculations.
Surfer~\cite{skarman2024Surfer} is a waveform viewer tightly coupled with the Spade language~\cite{skarmanSpadeHDLInspired2022conferencePaper}, supporting the representation of more abstract values of the language. 
Although it is built for extensibility, it does not yet offer native support for other HDLs, and it lacks the type information in the view.
Synopsys Verdi~\cite{VerdiAutomatedDebugwebpage} is a vendor waveform viewer with many advanced functionalities. 
Synopsys has demonstrated a new alpha version of the viewer and has cooperated with CIRCT to define the HGLDD debug format for supporting a more abstract view.
However, since HGLDD provides information only about FIRRTL as mentioned earlier, this alpha version provides a waveform visualization at a FIRRTL level rather than at the Chisel level.

Tywaves aims to provide a new kind of open source waveform visualization by showing source-level data types and offering support for Chisel.
Its integration within CIRCT would imply a new methodology for propagating information through an existing toolchain, and could contribute to the extension of support for other languages.

\section{Implementation}
\label{sec:implementation}
Section~\ref{sec:intro} and~\ref{sec:background} highlighted that the current flow of ChiselSim lacks inherent support for generating and propagating type information from Chisel. 
In addition, the debug information that CIRCT can generate is not available to testbench users.

In this section, the implementation of Tywaves is presented. 
The key idea is to preserve and propagate information during each transformation from one IR to another and make it available to a waveform UI. 
The functionality is integrated within the Chisel-CIRCT compilation pipeline. 
The specific updates consist of:
\begin{itemize}
    \item Updating Chisel elaboration to collect the type information and pass it to CIRCT through FIRRTL.
    \item Processing the new information and updating the debug dialect to handle the missing characteristics.
    \item Extending a waveform viewer to support Tywaves.
    \item Developing a ChiselSim API to connect all tools involved.
\end{itemize}
\subsection{Updates in Chisel}
During the first elaboration, Chisel gets translated to FIRRTL. Even though this IR keeps the structure and signal hierarchies, it cannot express any Scala abstraction or source type because this step executes all Scala meta-programming~code.

Chisel elaboration internally performs a sequence of transformations, called phases. Two of them are specifically responsible for the translation to FIRRTL: {elaborate} and {convert}. The former executes and elaborates the body of a module into a circuit hardware graph while the latter converts it into a FIRRTL equivalent graph, which is finally serialized and passed to the CIRCT compiler.

To enable Tywaves functionality, we have created a new phase in between, \texttt{AddTywavesAnnotation}, which executes only when a specific debug flag is set.
Hence, this addition does not affect the execution time of normal compilations when debug information is unnecessary, such as when the logic is compiled for synthesis. This is similar to the way the GCC compiler~\cite{GCCGNUCompilerwebpage} can run in debug mode when used with the GDB debugger~\cite{GDBGNUProjectwebpage} to provide useful information for catching errors in source code, in contrast to running GCC in release mode to create a faster and lighter executable.

This new phase parses each element of the Chisel circuit graph, extracts extra debug information, and outputs a new graph where each node is annotated with type information. The debug information is finally encoded in FIRRTL using the annotation mechanism, which allows the association of custom metadata to FIRRTL elements.
Table~\ref{tab:extracted-tywaves-info} summarizes the information collected by this new Chisel elaboration phase.

\begin{table}
    \centering
    \caption{Extracted Tywaves information using the new Chisel~elaboration phase in debug mode}
    \label{tab:extracted-tywaves-info}
\resizebox{\columnwidth}{!}{%
\begin{tabular}{ll|l}
\hline
\multicolumn{2}{c|}{\textbf{Information}} & \textbf{Example} \\ \hline
\multirow{2}{*}{Type name} & Class name & \texttt{Bool}, \texttt{SInt}, \texttt{MyBundle} \ldots  \\
                           & Chisel binding & \texttt{IO}, \texttt{Wire}, \texttt{Reg}, \texttt{OpResult} \ldots                  \\ && \\
\multirow{3}{*}{Parameter} &  Name                    &  \multirow{3}{*}{\texttt{class MyBundle(val \textbf{n}: \textbf{Int})}} \\
                           & Scala type              &                   \\
                           & Value                   &                   \\ &&\\
Enum definition            & Map $(int\rightarrow name)$ & $(0\rightarrow \texttt{IDLE}; 1 \rightarrow \texttt{A})$             \\ \hline   
\end{tabular}%
}
\end{table}

\subsection{Updated CIRCT debug dialect}
The 4 Ops in the current version of the CIRCT debug dialect misses some characteristics necessary to enable the full reconstruction of the Chisel view. 
None of them can represent Chisel type information, Scala parameters, definitions of enum types, specific debug information for modules, nor the type for subfields of aggregates.
The dialect is updated with new and modified operations to support these features of Chisel.
\subsubsection{New dbg.enumdef}
This new operation contains data to reconstruct the named variants of an enum type. Variants are internally implemented as a dictionary attribute: $(Int \rightarrow String)$ which can be converted to a hash map by a viewer to access variants with an $O(1)$ time complexity. It returns a result to associate the definition with other debug~Ops.

\subsubsection{Updated dbg.variable}
The operation has been updated with a new operand for the result of \textit{dbg.enumdef} and two new MLIR attributes which accommodate the type name and parameter info of Table~\ref{tab:extracted-tywaves-info}. The new attributes and operand are declared as optional entries to handle the case that the debug information is not generated during the Chisel elaboration.

\subsubsection{New dbg.subfield} The \textit{dbg.subfield} Op represents source information for subfields of aggregates to clearly separate them from variable declarations.
It has the same attributes and operands as \textit{dbg.variable} but it additionally returns a result that can be chained to debug aggregate Ops.
Although they may seem quite similar, having two distinct Ops for top declarations and subfields helps to {clearly distinguish the two roles}.

\subsubsection{New dbg.moduleinfo}
Types cover modules and instances as well as signals of a circuit. 
Hence, Tywaves defines \textit{dbg.moduleinfo} Op to store this information for a module and make it available in the compiler.
The Op is purely declarative and does not accept operands or return results.

\subsection{Integrating Tywaves in Surfer}
The new version of the debug dialect is serialized in an updated HGLDD file format to make the new information accessible from an external waveform viewer. 
This new format is emitted separately from the older one to avoid breaking other existing tools that use it.

Surfer~\cite{skarman2024Surfer} is a new open source waveform viewer, written in Rust, with an active community open to updates and extending support to other HDLs, including Chisel. 
It has been implemented with an emphasis on extensibility; the value rendering system is based on a \texttt{Translator} trait to create custom representations of the values from a simulation and ensure modularity for defining multiple translators.

\begin{figure*}[t]
    \centering
    \includegraphics[width=0.9\textwidth]{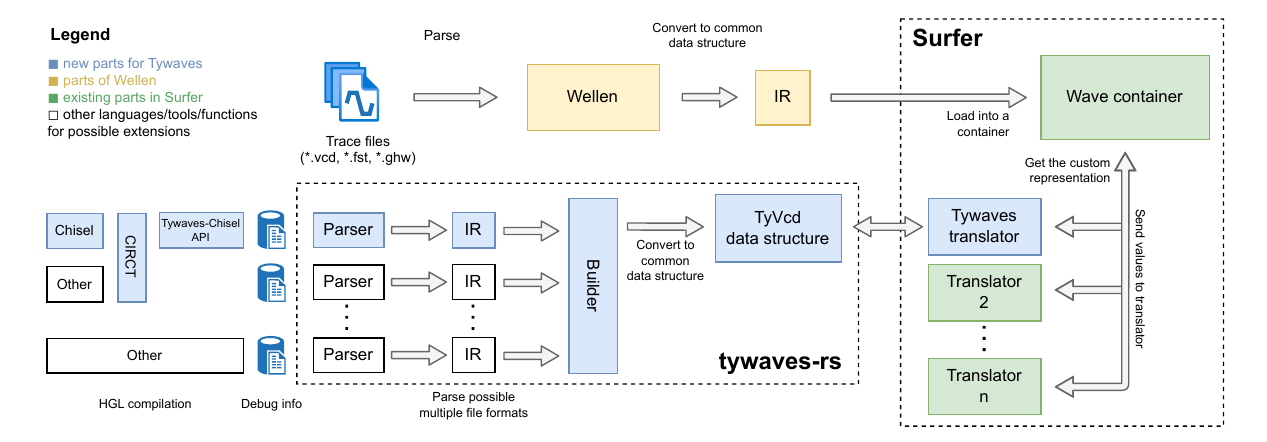}
    \caption{Tywaves integration within the Surfer waveform viewer}
    \label{fig:tywaves-surfer}
\end{figure*}

Figure~\ref{fig:tywaves-surfer} shows a high-level diagram of Surfer and how the Tywaves functionality is integrated.
The trace files are loaded and converted to an intermediate data structure by Wellen~\cite{KevinLaeufer2024Wellen}, providing a format-independent interface with the rest of the system.
Tywaves implements a specific Rust library, \texttt{tywaves-rs}, to bridge the debug information from CIRCT with the simulation values.
It follows the same logic as Wellen by converting an input format into a common intermediate data structure, \texttt{TyVcd}.
This allows for decoupling the internal functionality of the viewer from the reading debug interface and facilitates the extension to other languages, either using CIRCT or other compilers.
The \texttt{TywavesTranslator} communicates with \texttt{TyVcd} and processes the values such that signals are displayed with source-level types and hierarchies are preserved.

\subsection{Tywaves-Chisel API}
Tywaves-Chisel API provides two high-level simulators to improve the current functionality of ChiselSim and enable Tywaves directly from testbenches.

The state of the art of ChiselSim implements only a simple \texttt{EphemeralSimulator} for ephemeral simulations. 
It only enables simulating a circuit without keeping any information about the test, as the name suggests (e.g., no emission of a trace).
Thus, new functionalities are needed to enhance the user's experience:
\begin{itemize}
    \item \textbf{Parametric simulator}: parameterizes simulations through a set of predefined settings. For instance, \texttt{VcdTrace} to get the VCD trace file and \texttt{WithFirtoolArgs} to customize CIRCT compilation.
    \item \textbf{Tywaves simulator}: extends the previous simulator and calls the Chisel elaboration and CIRCT in order to generate and process Tywaves information. Additionally, it offers a setting to launch the new version of Surfer directly from a ChiselSim testbench.
\end{itemize}

\section{Use case: debugging ChiselWatt}
\label{sec:use-case}
To demonstrate the advantages of Tywaves over standard waveforms, we simulate and test ChiselWatt~\cite{blanchardAntonblanchardChiselwatt2024computerProgram}, a general-purpose soft-core processor in Chisel that implements a small version of the OpenPOWER ISA (Instruction Set Architecture)~\cite{OpenPOWERFoundationwebpage}.

\begin{figure}
    \centering    
  \subfloat[Tywaves output for ChiselWatt simulation\label{fig:chiselwatt-tywaves}]{
        \includegraphics[trim={0cm 0cm 20cm 0},clip, width=\linewidth]{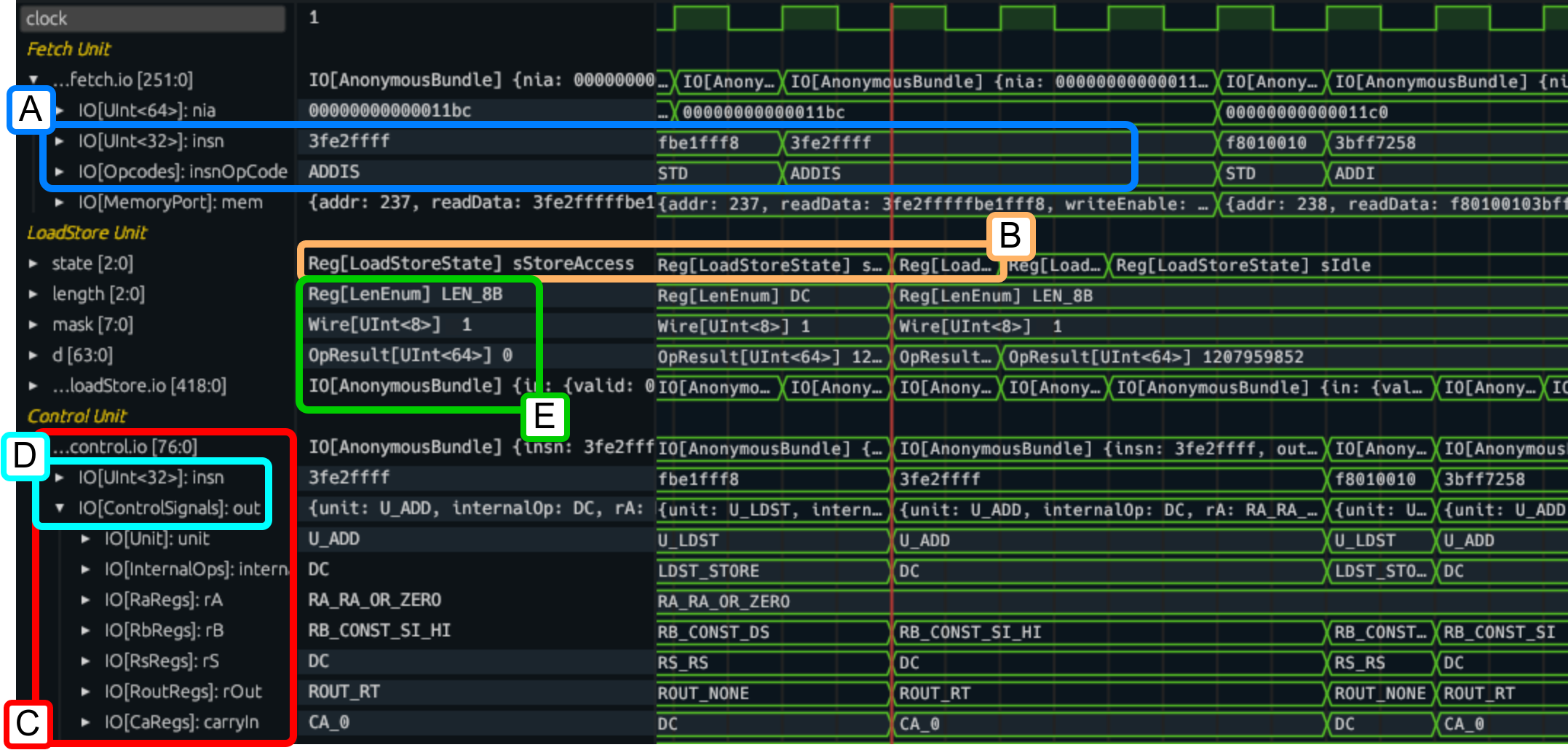}
    }
    \hfill
  \subfloat[Standard waveform output for ChiselWatt simulation   \label{fig:chiselwatt-vcd}]{
       \includegraphics[trim={0cm 0cm 20.5cm 0},clip, width=\linewidth]{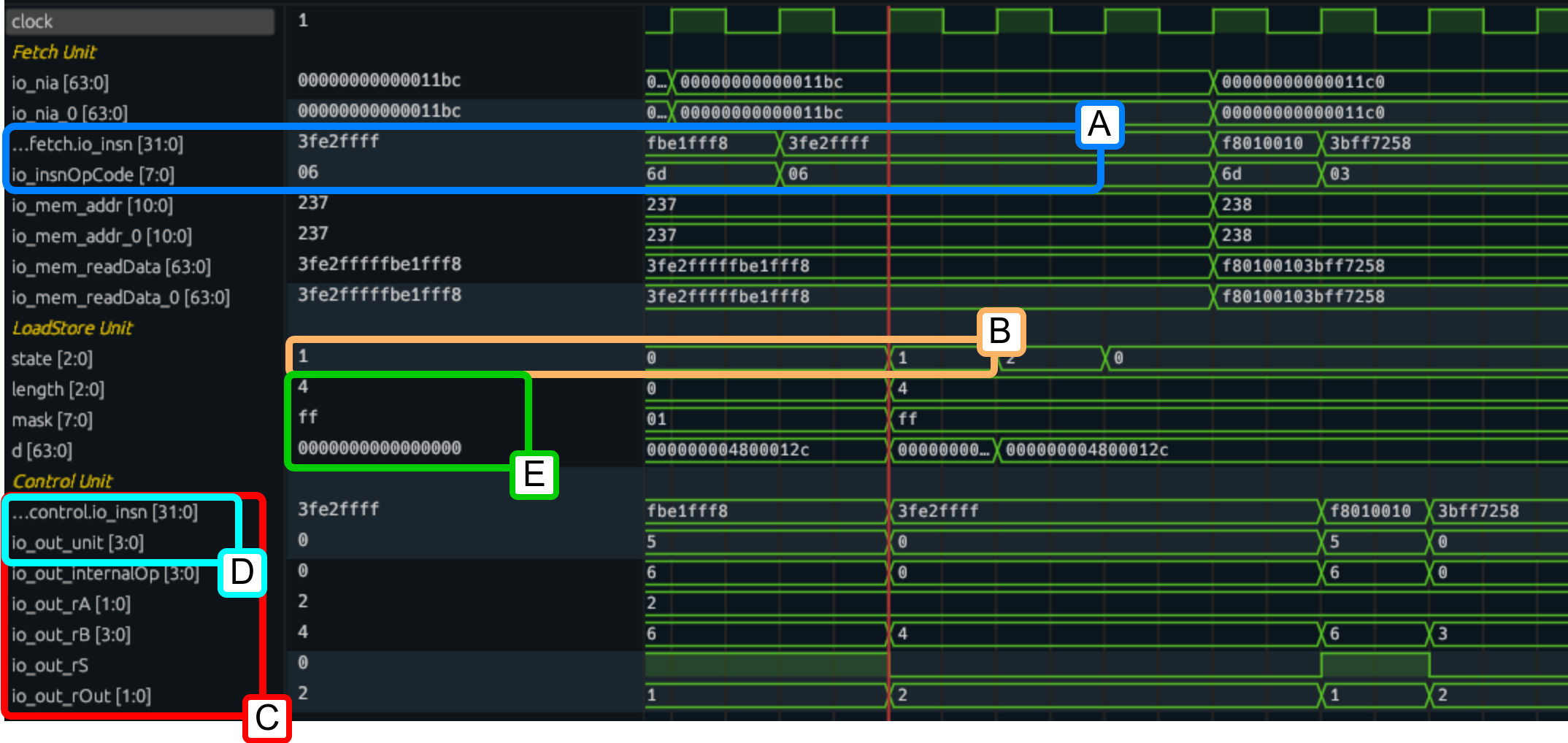}
    
   }
    \caption{Comparison between Tywaves and standard waveform outputs for simulation of ChiselWatt. \rounded{roundblue}{A} Opcode represented as enums; \rounded{roundorange}{B} State of computational units; \rounded{red}{C} Aggregated signal hierarchies; \rounded{roundseagreen}{D} Types enable clearer distinctions; \rounded{roundgreen}{E} Chisel bindings and Chisel-Scala types.}
    \label{fig:chiselwatt-waves}
\end{figure}

Figure~\ref{fig:chiselwatt-waves} compares the output waveforms from Tywaves with a standard view without Tywaves.
The instructions of an ISA might specify the operations to perform (\textit{opcodes}) as well as the \textit{operands} used for the computation.
In ChiselWatt, enumerations can be used to define opcodes, states of computational units, register identifiers, selectors, etc., allowing manipulation of named values rather than numbers.
For instance, in Figure~\ref{fig:chiselwatt-tywaves} it is easy to inspect which opcode 
is being fetched or the current state of a computational unit (Blocks A and B in the figure). 

Preserving the hierarchies of the aggregated signals allows us to keep the same original structure and variable names. 
In addition, this representation allows for a clear distinction of aggregates and fields from ground variables, as well as collapsing and loading signals of an aggregate as a whole data structure (Block C in the figure). 

The type information kept by Tywaves improves the understanding of the expected values and what a signal represents, e.g., distinguishing between instructions and control signals (Block D in the figure).
Furthermore, Tywaves preserves the distinction between Chisel bindings and Chisel-Scala types, enabling the identification between ports, wires, intermediate results, etc., (Block E in the figure).

As can be seen in Figure~\ref{fig:chiselwatt-vcd}, standard waveforms do not maintain the same level of abstraction.
Types and Chisel bindings are not shown, opcode and instruction values are shown as bytecode and aggregates might be flattened and represented as separate variables.
In Tywaves, this information is preserved to provide a more organized, understandable, and cleaner format that is closer to the code.

\section{Conclusions}
\label{sec:summary}
Using standard waveform viewers in modern HDLs is challenging, as designs expressed in classic HDLs are forced to include low-level details that are usually hidden from designers.
Tywaves addresses this issue by enabling source-level view and displaying custom data types for the Chisel language.
The compilation toolchain is updated to support generating and propagating extra information for debugging applications.
The use case presented to debug the ChiselWatt processor proves how designs implemented in Chisel benefit from a type-based waveform visualization compared to the standard tools targeting Verilog.

Although this paper presents an implementation for Chisel, Tywaves is built to be easily extended to other languages while keeping significant parts of the implementation unchanged. 
Leveraging standard frameworks and representations, such as CIRCT and MLIR, helps make Tywaves easy to extend to other HDLs.
The project is made open source on GitHub\footnote{Tywaves-Chisel GitHub: https://github.com/rameloni/tywaves-chisel-demo} to encourage the community to use it, and pull requests have been issued to Chisel, CIRCT, and Surfer for the updates presented.



\section*{Acknowledgment}
This research was performed with the support of the Eureka Xecs project TASTI (grant no.~2022005). The research benefited from conversations with the community and developers of Chisel, CIRCT, and Surfer on official channels and meetings.

\bibliographystyle{ieeetr}
\bibliography{references}
\end{document}